\documentclass[doublecol]{epl2} 
\usepackage{graphicx} 
\usepackage{amsmath} 
\usepackage{amssymb} 

\usepackage{soul}
\usepackage{stfloats}
\newcommand{\eqrefn}[1]{eq.~(\ref{#1})}
\newcommand{\figref}[1]{fig.~\ref{#1}}

\title{Work to insert a particle into an active fluid}

\author{Freddy A. Cisneros\inst{1} \and Alexandre Solon \inst{2}\and Jordan M. Horowitz \inst{3,4,5}\thanks{E-mail: jmhorow@umich.edu (corresponding author)}
}
\shortauthor{F. A. Cisneros \etal}

\institute{   
 \inst{1} Applied Physics Program, University of Michigan - Ann Arbor, MI 48109,
USA\\
 \inst{2} Sorbonne Universit\'{e}, CNRS, Laboratoire de Physique Th\'{e}orique de la Mati\`{e}re
Condens\'{e}e, LPTMC - F-75005 Paris, France\\
 \inst{3} Department of Biophysics, University of Michigan - Ann Arbor, MI 48109,
USA\\
\inst{4} Center for the Study of Complex Systems, University of Michigan - Ann
Arbor, MI 48109, USA\\
 \inst{5} Department of Physics, University of Michigan - Ann Arbor, MI 48109,
USA
}

\abstract{The chemical potential is defined as the work to
 quasi-statically add a particle to an equilibrium system. Inspired
 by this definition, we investigate how the work to add a particle to
 an active fluid depends on the activity, density, and insertion
 protocol. We find that the average work is protocol dependent and
 decreases with activity. Moreover, the work fluctuations retain
 asymmetric non-Gaussian tails even for slow particle insertions. We
 then compare the average particle-insertion work to the steady-state
 densities observed when two active fluids are brought into diffusive
 contact and observe opposing trends between density and work.}

\begin{document}

\maketitle

\section{Introduction}
A long-standing goal in the statistical physics community has been to
delineate the extent to which equilibrium thermodynamic principles can
be broadened and adapted to describe nonequilibrium behavior. For
example, in systematic treatments of thermodynamics, intensive state
parameters—like pressure, temperature, and chemical potential—are
defined as the work required to infinitesimally vary a conjugate
(extensive) coordinate~\cite{Callen}. A key prediction that follows
is that these quantities determine the conditions for equilibrium:
mechanical, thermal, diffusive, and chemical. It has long been
recognized that it would be valuable if analogous quantities could be
defined for nonequilibrium states.

New perspectives are coming into view with the emergence of active
matter: a class of nonequilibrium systems in which entities are driven
out of equilibrium by the individual consumption of
energy~\cite{Bowick2022,te_vrugt_metareview_2025}. This contrasts with more
traditional nonequilibrium systems that are driven by boundary
conditions or by the application of global (nonconservative)
forces~\cite{deGroot,Chou2011}. Active matter models exhibit a wide
range of emergent behaviors—including swarming, clustering, and
motility-induced phase separation—even when interactions are purely
repulsive~\cite{Cates2015,Bechinger2016,chate_dry_2020,alert_active_2022,granek_colloquium_2024}. These
features offer opportunities for engineering self-organizing
biomaterials~\cite{zkale2021} and designing targeted drug delivery
systems~\cite{Ghosh2020}, while also providing a framework to explore
the role of nonequilibrium processes in biological phenomena like cell
migration and bacterial motility~\cite{Vicsek2012,Banerjee2019}.

Research into the intensive parameter pressure in active systems has
been especially fruitful due to its unambiguous mechanical definition
as the force per unit area applied to a boundary. Studies of this
force have shown that mechanical pressure depends on the specific
details of boundary interactions~\cite{Solon2015}. As a result,
pressure generally fails to satisfy an equation of state involving
only bulk properties. Nevertheless, under simplifying assumptions
(that we will make in this Letter), in particular the absence of
alignment interactions, the mechanical pressure does obey an equation
of state~\cite{fily_mechanical_2017} and equalizes between two
coexisting
phases~\cite{solon_pressure_2015,solon_generalized_2018,Solon2018}. Such
systems thus seem to be the best testing ground for extending
thermodynamic concepts to active systems.

Unlike pressure, the chemical potential lacks a mechanical definition.
Gibbs introduced the chemical potential in 1876 as the adiabatic work
required to insert a particle into a substance at constant
pressure~\cite{Gibbs1876}. It is then a consequence of thermodynamic
principles that this energetically-defined quantity also predicts a
variety of equilibrium properties. That is, many distinct experimental
measurements yield the same value, namely, the chemical potential.
Out of equilibrium, these different procedures are expected to give
different results, but each may still serve as a provisional
operational definition of a nonequilibrium chemical potential. One
such approach mirrors the equilibrium role of chemical potential in
diffusive equilibrium: particles flow from regions of higher to lower
chemical potential until the chemical potentials are equal
\cite{Callen}. This motivates defining a nonequilibrium chemical
potential as a parameter that satisfies a similar transitive property
across steady states. Simulations of externally-driven lattice gases
suggest that such a definition can hold for certain
cases~\cite{Pradhan2010,Pradhan2011}. Another perspective comes from
linear irreversible thermodynamics, which postulates that the chemical
potential gradient governs not only the direction but also the rate of
particle flow through a linear constitutive relation~\cite{deGroot}.
The authors of refs.~\cite{Takatori2015,paliwal_chemical_2018} connect
this definition to mechanical properties of an active fluid, allowing
for predictions of phase coexistence. A third approach relates
chemical potential to fluctuations in particle number, as encoded in a
large deviation function or free energy. When the macroscopic dynamics
satisfy certain factorization properties, this leads to an effective
chemical potential that accurately predicts steady-state
distributions~\cite{Bertin2006,Bertin2007,Pradhan2010,Guioth2018,Guioth2019,Guioth2019b}.
While these approaches stem from different features of the equilibrium
definition, a common thread is that the dynamics of particle exchange
across interfaces influence steady-state behavior in nonuniversal
ways—obscuring the possibility of defining a nonequilibrium chemical
potential using only bulk properties.

Although particle-insertion work is a well-established method for
computing the equilibrium chemical potential~\cite{Vaikuntanathan2011,
 Widom1963}, to the best of our knowledge, this energetic definition
has not been explored as a potential nonequilibrium generalization.
Motivated by Gibbs’ original formulation, we computationally
investigate how the work to add a particle to an interacting active
fluid depends on activity, density, and insertion protocol. We find
that the particle-insertion work exhibits larger fluctuations in
active fluids compared to their equilibrium counterparts and that the
work depends quantitatively on the protocol used to add the particle,
even in slow insertions.

We then ask whether the density dependence of this work can predict
diffusive equilibrium between two coupled active fluids—one
interacting and the other not. In other words, can particle-insertion
work serve as a useful nonequilibrium definition of chemical
potential? Our results show that the insertion work measured in the
bulk of each fluid can differ, a discrepancy that we trace to strong
variations in particle density and activity near the interface.

\section{Setup}
Work measurements were performed in simulations of $N$ interacting
active Brownian particles (ABPs) in a two-dimensional system of
horizontal length $L_x$ and vertical length $L_y$ with periodic
boundary conditions. The time-dependent particle positions
${\boldsymbol x}_i$ and self-propulsion angles $\theta_i$ relative to
the $+x$-axis, $i=1,\dots,N$, evolve according to the coupled
overdamped Langevin equations with unit friction coefficient,\noindent
\begin{subequations} 
\label{eq:ABP}
\begin{align}
\dot{\boldsymbol{x}}_i &= v\hat{\boldsymbol{u}}_i + \sqrt{2D}\boldsymbol{\xi}_i - \sum_{j\neq i} \nabla_i U(|{\boldsymbol x}_i-{\boldsymbol x}_j|), \label{eq:Leq}\\ 
\hat{\boldsymbol{u}}_i &= (\cos\theta_i, \sin\theta_i),\\
\dot{\theta}_i &= \sqrt{2D_R}\eta_i. 
\end{align}
\end{subequations}
Thermal fluctuations from the surrounding fluid are modeled as
independent Gaussian white noises $\boldsymbol{\xi}_i$ with
translational diffusion coefficient $D$. Self-propulsion has fixed
magnitude $v$, and its direction $\hat{\boldsymbol{u}}_i$ diffuses due
to independent Gaussian white noises $\eta_i$ with rotational
diffusion coefficient $D_R$.

Particles interact via a pairwise, short-range harmonic potential that
depends only on the interparticle distance
$r_{ij} = |{\boldsymbol x}_i - {\boldsymbol x}_j|$,
\begin{equation}
U(r_{ij}) = \frac{k}{2}(\sigma - r_{ij})^2 \Theta(\sigma - r_{ij}), \label{eq:Potential}
\end{equation}
with interaction strength $k$, interaction length $\sigma$, and the
Heaviside step function $\Theta$ ensures that interactions vanish when $r_{ij} \geq \sigma$.

Within this framework, ABPs can be interpreted as soft disks of diameter $\sigma$ with packing fraction $\phi$ and number density $\rho$:
\begin{equation}
\phi = \frac{N\pi \sigma^2}{4L_x L_y} = \frac{\rho \pi\sigma^2}{4},
\label{eq:packing}
\end{equation}
where $L_x L_y$ is the area of the system.
Their motion is
 parameterized by two dimensionless numbers: (i) The effective
 diffusion coefficient $D^\text{eff} = D + v^2/(2D_R)$, which
 characterizes the motion of an isolated particle on long time-scales
 compared to the persistence time $1/D_R$ and (ii) The Peclet number
\begin{equation} 
\text{Pe} \equiv \frac{3v}{\sigma D_R} = \frac{3\ell}{\sigma}, \label{eq:peclet}
\end{equation}
with $\ell\equiv v/D_R$ the persistence length, which characterizes
the persistence of the particles. To minimize finite-size effects, we
ensured that $L_x, L_y \gg \ell$.

In this Letter, we will consider the purely active case, active
particles (AP) with $v\neq 0$ and $D=0$, and compare to Brownian
particles (BP) with $v=0$ and $D\neq 0$. To have a meaningful
comparison, we fix $D^\textrm{eff} = 3.33$ for both types of
particles. We then vary the Peclet number of the AP. In the low Peclet
limit $\text{Pe} \ll 1$, the particles reorient on a short time scale
compared to that of collisions and one then expects to recover the
behavior of BP with the same diffusion coefficient
$D=D^\textrm{eff}$~\cite{solon_active_2015,Bechinger2016}.

\section{Particle-Insertion Work}
To measure the work required to add a particle, we evolved a fluid of
$N$ ABPs according to \eqrefn{eq:ABP} until a steady state was
reached. At this point, we randomly added a single non-interacting
probe particle, that is otherwise identical to the fluid
particles. For reference, this moment is defined as $t=0$. The
interaction potential between the probe and the surrounding fluid
particles was then switched on deterministically over a fixed interval
$\tau$, after which the probe becomes indistinguishable from the other
particles. Thus, the interaction potential is explicitly
time-dependent, $U_p(r_{pj},t)$, where
$r_{pj} = |{\boldsymbol x}_p - {\boldsymbol x}_j|$ is the distance
between the probe particle and a fluid particle.

We consider two distinct protocols to ramp up the interaction between the probe and fluid particles depending on whether we manipulate the interaction strength or the interaction length in eq. (2):
\begin{itemize}
\item \textbf{$k$-Protocol:} The interaction strength increases linearly from $0$ to $k$,
\begin{equation}
k_p(t) = k\frac{t}{\tau},
\end{equation}
while the interaction length remains fixed, $\sigma_p(t) = \sigma$.
\item \textbf{$\sigma$-Protocol:} The interaction length increases linearly from $0$ to $\sigma$,
\begin{equation}
\sigma_p(t) = \sigma\,\frac{t}{\tau},
\end{equation}
while the interaction strength is fixed, $k_p(t) = k$.
\end{itemize}
In both protocols, the probe particle initially has no interaction with its neighbors ($U_p(r_{pj},0)=0$ at $t = 0$), but at the end of the process the probe particle's interactions match those of the other particles ($U_p(r_{pj},\tau)=U(r_{pj})$ at $t = \tau$).

Due to thermal and active fluctuations, the work done during the insertion depends on how the distance $r_{pj}(t) = |{\boldsymbol x}_p - {\boldsymbol x}_j|$ evolves in time. Specifically, for each protocol, the work in each realization of the dynamics is determined by integrating over the system's dynamical trajectory as
\begin{equation}
W = \int_0^\tau \sum_{j \neq p} \partial_t U_p(r_{pj}(t), t) \, dt.
\end{equation}
In general, the average and fluctuations of $W$ depend on the protocol
and its duration. However, when this insertion process is carried out
adiabatically slowly for the equilibrium Brownian particle dynamics,
(i) the average work should be independent of the protocol (path) \cite{Callen}, and
(ii) fluctuations are expected to be Gaussian, at least below a
duration-dependent critical work
value~\cite{Speck2004,Hoppenau2013}. Throughout, measurements were
performed by simulating the above dynamics using the Euler algorithm
with a time step $\Delta t = 10^{-4}$ on a system of size
$40 \times 40$, relaxed to steady-state for $t_\text{ss} = 10^3$.

\begin{figure}[h!]
 \centering
 \includegraphics[width=0.77\columnwidth]{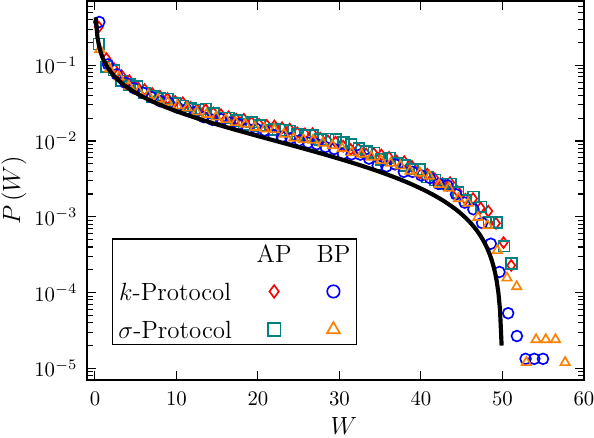}
 \caption{Instantaneous particle-insertion work distribution $P(W)$ in a dilute system. The theoretical distribution in \eqrefn{eq:dist} (solid line) is shown for $W \in (0, k\sigma^2/2)$, alongside histograms of measured values (markers). A total of $6 \times 10^4$ work measurements were collected for each combination of dynamics, system type—Brownian particle (BP) or active particle (AP)—and protocol. Simulation parameters: $\tau = 10^{-3}$, $\phi = 5 \times 10^{-2}$, $\sigma = 1$, $k = 100$, $\text{Pe} = 20$, and $D^\text{eff} = D = 3.33$.}
 \label{fig:winst}
\end{figure}

We first computed the particle-insertion work for infinitely fast
protocols ($\tau \to 0$) in a dilute system, where analytical progress
is possible. Computational results for the work distribution $P(W)$
for both protocols in both the active and Brownian particle dynamics
are presented in \figref{fig:winst}.
The work distribution is independent of the protocol and the
dynamics. In this limit, work is only done when, at the instant the
probe particle is added, it overlaps with a fluid particle. Moreover,
this work depends solely on this distance $r \le \sigma$:
\begin{equation}\label{eq:workInstant}
W \xrightarrow{\tau \to 0} \frac{k}{2} (\sigma - r)^2.
\end{equation}
Since the fluid is dilute, we only consider the chance of overlap with
a single fluid particle in calculating the probability. Namely, the
probability that the probe particle is added at a distance $r$ around
a fluid particle given that $r \le \sigma$ is
$P_\sigma(r) = 2 r / \sigma^2$. The work distribution then follows
from a change of variables, $P(W) = P_\sigma(r) |dr/dW|$, using
$r = \sigma - \sqrt{2W/k}$ from \eqrefn{eq:workInstant},
\begin{equation}
P(W) = \frac{2}{\sigma^2 k} \left( \sqrt{\frac{k \sigma^2}{2W}} - 1 \right).
\label{eq:dist}
\end{equation}
This prediction, shown by the black line in \figref{fig:winst}, agrees well with the computational measurements up to the theoretical maximum value $k\sigma^{2}/2$. Work values exceeding this bound are likely attributable to rare instances in which the probe particle interacts with more than one fluid particle, or to the fact that the switching time is finite, $\tau = 10^{-3}$, rather than strictly zero. During this finite interval, particles can undergo small displacements, which occasionally increases their overlap relative to the instantaneous configuration assumed in the theoretical derivation.

\begin{figure*}[!b]
 \centering
 \includegraphics[width=0.8\textwidth]{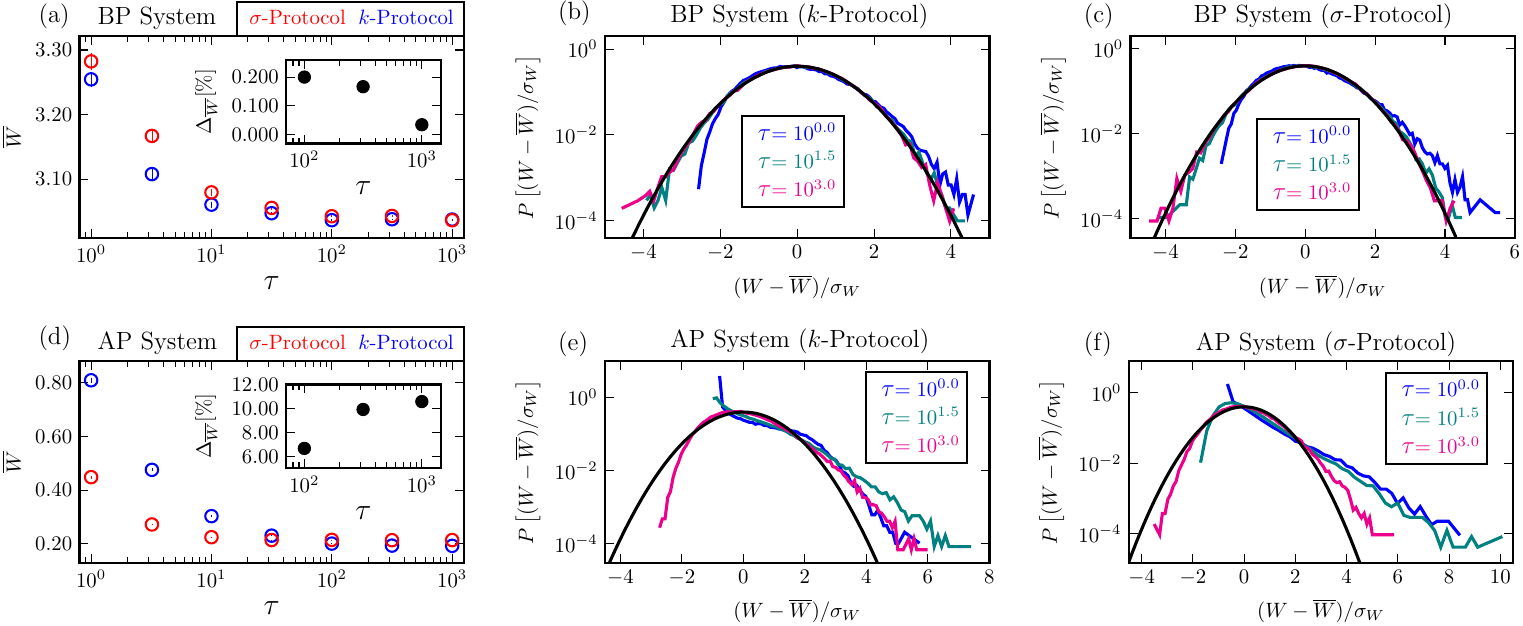}
 \caption{Particle-insertion work statistics for Brownian particles
 (BP, top row) and active particles (AP, bottom row) at finite
 switching times $\tau$. (a), (d) Average work when the interaction length ($\sigma$-protocol) and interaction strength
 ($k$-protocol) are varied. (b), (e) Standardized work
 distributions for the $k$-protocol, compared to the normal
 distribution (black lines). (c), (f) Standardized work distributions
 for the $\sigma$-protocol, compared to the normal distribution (black
 lines). Statistics in (a), (d) were computed from $10^3$ measurements,
 while $10^5$ measurements were used for the standardized work
 distributions in (b), (c), (e), (f). Simulation parameters: $\phi = 0.3$,
 $\sigma = 1$, $k = 100$, $\text{Pe} = 20$, $D^\text{eff} = D = 3.33$.}
 \label{fig:workdist}
\end{figure*}

For finite protocol durations and denser systems, we resort to the
computational analysis presented in \figref{fig:workdist}, where we
measured the particle-insertion work distribution for both protocols
and dynamics as a function of protocol duration $\tau$.
In the fluid of equilibrium Brownian particles, the average work
$\overline{W}$ to insert the probe particle for the two protocols
coalesces in the quasistatic limit, $\tau \gg 1$
(\figref{fig:workdist}(a)). The inset shows that the relative work
difference between the two protocols,
\begin{equation} 
\Delta_{\overline{W}}(\tau) = 2 \left| \frac{\overline{W}_{\sigma}(\tau) - \overline{W}_{k}(\tau)}{\overline{W}_{\sigma}(\tau) + \overline{W}_{k}(\tau)} \right| \times 100\%,
\label{eq:reldiff} 
\end{equation}
is only $\Delta_{\overline{W}} = 0.035\%$ at $\tau = 10^3$. This
aligns with expectations from equilibrium thermodynamics that the work
performed during a quasistatic process between two equilibrium
states—here, transitioning from a system with $N$ to $N+1$ interacting
particles—should be independent of the protocol. Moreover,
 equilibrium theory predicts that small work fluctuations approach a
 Gaussian distribution as the protocol duration
 increases~\cite{Speck2004}, and, as expected, the standardized work
 distributions in \figref{fig:workdist}(b) and (c)—centered and rescaled by
 their variance $\sigma_W$—exhibit progressively reduced deviations
 from a quadratic form on a log-scale when $\tau$ increases. At our
 longest insertion time $\tau = 10^3$, non-Gaussian
 tails~\cite{Hoppenau2013} are fully suppressed within the precision
 of our measurement. Together, these results demonstrate that our
 simulation method and work-measurement protocol reproduce
 equilibrium thermodynamic expectations, providing a reliable
 baseline for comparison with the active-particle dynamics.

For active particle dynamics, the average work $\overline{W}$ for the
two insertion protocols differs, as shown in \figref{fig:workdist}(d). Even for long insertion durations, the difference remains as large as $\Delta_{\overline{W}} = 10.6\%$ at
$\tau = 10^3$. Although expected, it is worthwhile to confirm this in
the current scenario. The work distributions exhibit pronounced
non-Gaussian tails in \figref{fig:workdist}(e) and (f), even for the
longest insertion time tested. We speculate that this
 increase in large, rare positive work fluctuations arises from
 persistence in the active particle motion: when two self-propelling
 particles interact with opposing propulsion directions, they can
 remain in contact for relatively long times of order $1/D_R$. This
extended interaction time increases the integrated force during the
interaction, generating infrequent high-work events.

\begin{figure}[tb]
 \centering
\includegraphics[width=0.73\columnwidth]{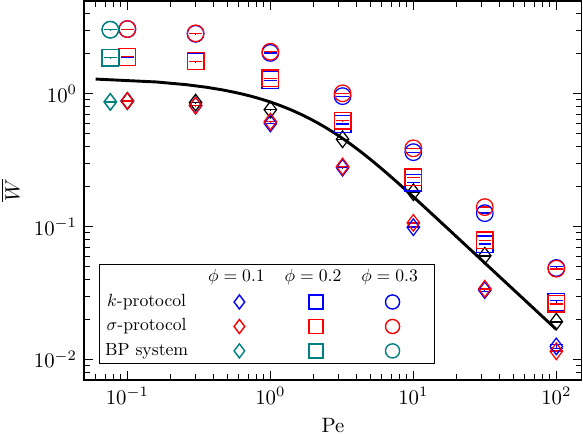}
 \caption{Mean particle-insertion work $\overline{W}$ as a function of Peclet number $\text{Pe}$ and packing fraction $\phi$  for the two protocols. Simulations of BP (green markers) are
 shown at an arbitrary $\text{Pe}$ on the left to compare with the
 small Peclet limit of AP. Solid black lines show the mean-field
 prediction $W_{\rm MF}$ from eq.~(\ref{eq:West}) for
 $\phi = 0.1$. The black diamonds show the mean insertion work for
 a $\sigma$-protocol at $\phi=0.1$ in which the added particle is
 held fixed. All data points are computed from $3 \times 10^{3}$
 independent realizations, giving statistical error bars that are
 smaller than the symbol sizes. Parameters: $\tau = 10^{3}$,
 $\sigma = 1$, $k = 100$, and $D^{\text{eff}} = D = 3.33$.}
 \label{fig:wvspe}
\end{figure}

Lastly, in \figref{fig:wvspe}, we examine the dependence of the
average work $\overline{W}$ on activity through the Peclet number Pe,
at fixed effective diffusion coefficient $D^\textrm{eff}$, for various
packing fractions $\phi$. For $\text{Pe} \ll 1$, dynamics are
diffusion dominated and behave similarly to Brownian particles. This
is supported by the agreement of the active particle average work with
the Brownian particle simulations at $D = D^\text{eff}$ for low Pe
($\text{Pe} = 10^{-1}$). The average work $\overline{W}$ also
increases with packing fraction as the probe particle interacts with
more fluid particles in denser systems. Beyond this increase
 in the mean, higher packing fractions seem to slightly reduce the
 relative fluctuations $\overline{W}/\sigma_W$, at least in the
 gaseous regime we explored (not shown), which may be related to
 the fact that the system becomes more homogeneous.

Perhaps surprisingly, the average work decreases with increasing
activity (larger Pe) for each packing fraction and
protocol. To rationalize this behavior, let us provide a
 simple mean-field estimation of the work needed to insert a
 non-moving particle (this constitutes a mean-field approximation
 since this removes correlations between the positions of
 the probe particle and the bath particles). We further work in the
 limit of quasi-static $\sigma$-protocol, a dilute bath and hard-core
 interaction potential $k \to \infty$. Under these assumptions, the
 insertion work is directly given as the work needed to counteract
 the mechanical pressure $P(r)$ ({\it i.e.} the mean force per unit area) exerted by a bath of non-interacting AP on a probe of size
 $r=\sigma_P(t)$
 \begin{equation}
 \label{eq:West}
 W_{\rm MF}=\int_0^\sigma 2\pi r P(r) dr.
 \end{equation}
 
 The pressure $P(r)$ depends on the radius of the probe since
 the curvature affects the residence time of the bath particles~\cite{fily_dynamics_2014,nikola_active_2016}. Although, we
 do not have an analytical expression for $P(r)$, a phenomenological
 expression has been built in ref.~\cite{smallenburg_swim_2015} from
 numerical measurements. Using this expression gives $W_{\rm MF}$ as
 shown in~\figref{fig:wvspe} for $\phi=0.1$ (full black
 line).

 Despite the approximations, $W_{\rm MF}$ gives the correct order of
 magnitude for the insertion work and the correct trends at small and
 large Peclet number. For $\text{Pe}\ll 1$, $P=D^\text{eff}\rho$
 independent of $r$, so that~(\eqref{eq:West}) gives
 $W_{\rm MF}=4 D^\text{eff}\phi$. For $\text{Pe}\gg 1$
 $P(r)=\alpha D^\text{eff}\rho r/\ell$ with $\alpha\approx 0.63$
 estimated numerically in~\cite{smallenburg_swim_2015}, which using~\eqrefn{eq:West} gives
 $W_{\rm MF}=8\alpha D^\text{eff}\phi/\text{Pe}$. This explains the
 decrease in insertion work when $\text{Pe}$ increases, which comes from
 the decrease in the pressure exerted by the bath.

 We find that the discrepancy between the insertion work computed in
 microscopic simulations and the mean-field estimation is due, for
 the most part, (i) At high $\text{Pe}$ to the mean-field assumption
 that the added particle is immobile. Repeating the microscopic
 measurement with an immobile probe (black diamonds in~\figref{fig:wvspe}) indeed gives values much closer to
 $W_{\rm MF}$. (ii) At low $\text{Pe}$ to the assumption of hard
 potential which breaks down since $v=6 D^\textrm{eff}/\text{Pe}$
 becomes large, and thus the particles effectively softer. We found that
 increasing $k$ improves the agreement with the theory in this limit
 (not shown).

\section{Excess Chemical Potential}
Next, we investigate the extent to which particle-insertion work can
be used to predict properties of diffusive equilibrium in active
systems. To this end, recall that in equilibrium systems the chemical
potential is comprised of an ideal and an excess
contribution~\cite{Frenkel2023}:
\begin{equation}\label{eq:muDef}
\mu = \mu^{\rm id} + \mu^{\rm ex}.
\end{equation}
The ideal part in two dimensions,
$\mu^{\rm id} = kT \log(\rho\lambda_{\rm th}^2)$, with
$\lambda_{\rm th}$ the thermal de Broglie wavelength, captures the
work to quasistatically increase the size of phase space when a probe
particle is added to the system. The excess contribution
$\mu^{\rm ex}$ arises from the mechanical work required to
quasistatically turn on interactions with the probe particle $U_p$,
precisely the work we analyzed above. This formulation is particularly
useful when two equilibrium systems, $A$ and $B$, are brought into
diffusive contact. Particle exchange proceeds on average until their
chemical potentials equalize, $\mu_A = \mu_B$. Via \eqrefn{eq:muDef},
this condition leads to a prediction for the densities of the two
systems:
\begin{equation}\label{eq:muEq}
\mu^{\rm ex}_A - \mu^{\rm ex}_B = kT \log(\rho_B / \rho_A).
\end{equation}
Lacking a clear definition of chemical potential in active systems, we
interpret the quantities in \eqrefn{eq:muEq} in a way that extends to
nonequilibrium settings. Specifically, we identify $\mu^{\rm ex}$ with
the particle-insertion work, as defined in the previous sections, and
replace the thermal energy $kT$ with the effective diffusive noise
$D^{\rm eff}$. We then ask: if two active systems are brought into
contact and allowed to exchange particles, how is the
particle-insertion work related to their bulk densities?

In this study, we take one system to be an interacting active fluid
and the other a non-interacting (ideal) active gas, which serves as a
natural reference for defining a nonequilibrium chemical
potential. Ideally, particles would diffuse freely between the two
systems through an interface. However, even in equilibrium Brownian
dynamics, such particle exchange would violate detailed balance, as
work must be performed to turn interactions on or off when a particle
crosses the interface~\cite{Frenkel2023}. While equilibrium
simulations can enforce this using an acceptance rule (e.g., the
Metropolis algorithm), no principled method exists to do so in active
systems. To circumvent this issue, we introduce a finite-width
interpolation region in which the interaction strength is smoothly
modulated. This construction avoids abrupt force
 discontinuities at the interface and permits steady particle
 exchange without imposing an explicit acceptance rule, thereby
 allowing to compare the chemical potentials computed from
 particle-insertion work and from the coexisting bulk densities.

\begin{figure}[h!]
 \centering \includegraphics[width=0.82\linewidth]{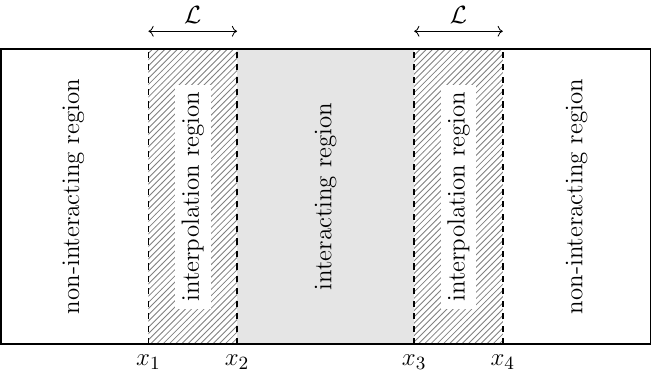}
 \caption{Schematic of the system partitioning. The system is
 divided into an interacting region, a non-interacting region of
 same size, and two interpolation regions each of width of
 $\mathcal{L}$.}
 \label{fig:system}
\end{figure}

Specifically, we simulate a periodic system partitioned into four
regions, as shown in \figref{fig:system}: an interacting region, a
non-interacting region, and two interpolation regions of width
$\mathcal{L}$.
The interaction potential in these regions
\begin{equation}
\mathcal{U}(r_{ij}, d_{ij}) = U(r_{ij}) \mathcal{F}(d_{ij}),
\label{eq:parpoten}
\end{equation}
smoothly varies the magnitude of $U(r_{ij})$ in \eqrefn{eq:Potential}
through a modulation function that depends on the center-of-mass
distance $d_{ij} = (x_i + x_j)/2$ along the $x$-axis,
\begin{equation}
\mathcal{F}(d_{ij}) = 
\begin{cases}
 \frac{d_{ij} - x_1}{\mathcal{L}}, & x_1 \leq d_{ij} < x_2, \\
 1, & x_2 \leq d_{ij} < x_3, \\
 \frac{x_4 - d_{ij}}{\mathcal{L}}, & x_3 \leq d_{ij} < x_4, \\
 0, & \text{otherwise}.
\end{cases}
\label{eq:indfunc}
\end{equation}
This produces a linear interpolation between interacting and
non-interacting behavior. The interacting and non-interacting regions
are each $100\sigma \times 100\sigma$, while the interpolation regions
are $\sigma \times 100\sigma$. The system is relaxed to steady state
over a duration $t_\text{ss} = 10^3$ and evolved using Euler
integration with time step $\Delta t = 10^{-4}$. Bulk densities (far
from the interface) are sampled every $10^3$ time step. This setup can
be seen as a practical way to build a grand-canonical ensemble out of
equilibrium. However, compared to other
proposals~\cite{sasa_steady_2006,guioth_nonequilibrium_2021}, we do
not assume slow exchange dynamics between the different
regions.
\begin{figure*}[!b]
 \centering
 \includegraphics[width=0.87\textwidth]{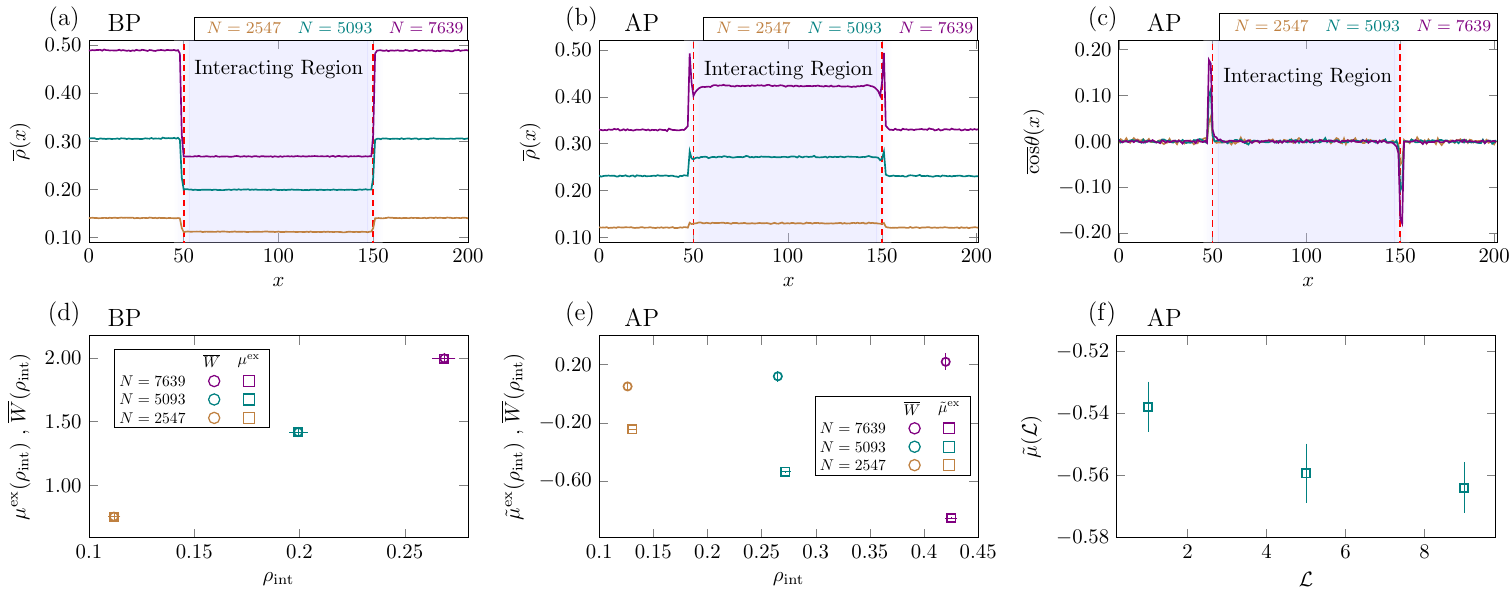}
 \caption{Comparison of particle insertion work and excess chemical
 potential: Vertical dashed lines in panels (a)–(c) indicate the
 interpolation regions. (a) Particle density $\rho(x)$ for the
 Brownian particle (BP) dynamics. (b) Particle density $\rho(x)$
 for the active particle (AP) dynamics. (c) Polarity profile for
 the AP system. (d) Excess chemical potential
 $\mu^{\rm ex}(\rho_{\rm int})$ and average work
 ${\overline W}(\rho_{\rm int})$ for Brownian particle dynamics. (e)
 Average work ${\overline{ W}}(\rho_{\rm int})$ and
 $\tilde\mu^{\rm ex}(\rho_{\rm int})$ for the AP system under the
 $\sigma$ protocol. (f) Dependence of $\tilde\mu^{\rm ex}$ on
 interpolation width $\mathcal{L}$ for the case where $N =
 5093$. Parameters: $\tau = 10^3$, $\sigma = 1$, $k = 100$,
 $v = 1$, $D^{\rm eff} = D = 3.33$.}
 \label{fig:chempot}
\end{figure*}

For equilibrium Brownian dynamics, the average density
$\overline{\rho}(x)$ is shown in \figref{fig:chempot}(a) as a function
of particle number $N$. We compared the excess chemical potential,
$\mu^{\rm ex}(\rho_{\rm int})$, computed using the average bulk
densities in the interacting ($\rho_{\rm int}$) and non-interacting
($\rho_{\rm non-int}$) regions, to the average particle-insertion work
$\overline{W}(\rho_{\rm int})$. Since no work is required in the
non-interacting region, the two agree as expected from
\eqrefn{eq:muEq}, as seen in \figref{fig:chempot}(d).

For active dynamics, the average density $\overline{\rho}(x)$ and
polarity $\overline{\cos}\theta(x)$ are shown in
\figref{fig:chempot}(b) and (c). In analogy to the equilibrium case, we
compute
\begin{equation}
\tilde{\mu}^\textrm{ex}(\rho_\text{int}) = D^{\rm eff} \log \left( \frac{\rho_{\rm non-int}}{\rho_{\rm int}} \right)
\end{equation}
and compare it to $\overline{W}(\rho_{\rm int})$ in
\figref{fig:chempot}(e). These two quantities not only differ, but
also exhibit opposite trends with increasing density: $\rho_{\rm int}$
increases relative to $\rho_{\rm non-int}$, leading to a decrease in
$\tilde{\mu}^\textrm{ex}(\rho_\text{int})$, while $\overline{W}$
increases. This strikingly different behavior must arise from the
pronounced interfacial structure since for active systems
 the details of the interfacial profile are known to affect the
 coexisting densities~\cite{Solon2018,Mahault2023}. Here, density
 spikes at the interface (\figref{fig:chempot}(b)) and the
 preferential orientation of those particles into the interacting
 region (\figref{fig:chempot}(c)) suggest that interactions slow
 entry into the interacting region while persistence prevents
 particles from diffusing away, leading to local accumulation. In
 principle, one could design an alternative scheme to reduce or
 remove the interfacial spikes, e.g. by imposing a special acceptance
 protocol at the boundary. However, at the moment, there is no
 theoretical consideration to guide us in designing such a protocol
 and we thus consider only the interface of~\figref{fig:system}.
Finally, \figref{fig:chempot}(f) demonstrates that the average
densities depend on the interpolation length $\mathcal{L}$, further
underscoring the role of interface interactions in shaping
steady-state properties in active systems.

\section{Summary} We analytically and computationally computed the
work to insert a particle into an active fluid. Unlike in the
equilibrium setting, the distribution of work fluctuations retains
asymmetric non-Gaussian tails even for insertion durations for which
the equilibrium work distribution appears Gaussian at all observed
work values. Additionally, the average particle-insertion work in the
active fluid decreases with increasing activity.

We then compared the particle-insertion work to the steady-state
densities obtained when two different active fluids were brought into
diffusive contact. Here, activity led to very different densities and
activity orientations in the bulk \textit{vs.} the interface, challenging
any connection with the particle-insertion work in the bulk. This
observation aligns with the consensus in the literature that in active
fluids the behavior of the fluid at boundaries and interfaces can
differ dramatically from the bulk and any thermodynamic framework
would need to systematically incorporate these effects.

\acknowledgments This work was partially supported by a grant from the
Thomas Jefferson Fund, a program of FACE. FAC and JMH acknowledge
financial support from the the National Science Foundation under Grant
No.\ 2142466 and the Chan Zuckerberg Initiative under Grant No.\
2024-350560. FAC acknowledges financial support from the Fulbright
U.S. Student Program, sponsored by the U.S. Department of State and
the Franco-American Fulbright Commission. The content is solely the
responsibility of the authors and does not necessarily represent the
official views of the Fulbright Program, the U.S. Government, or the
Franco-American Fulbright Commission.\newline

\textit{Data availability statement}: The data that support the findings of this study are available upon reasonable request from the authors.

\bibliographystyle{eplbib.bst}
\bibliography{export.bib}

\end{document}